\documentstyle[12pt]{article}

\def\thefootnote{\fnsymbol{footnote}}

\newcommand{\eq}{\begin{equation}}
\newcommand{\en}{\end{equation}}
\newcommand{\eqa}{\begin{eqnarray}}
\newcommand{\ena}{\end{eqnarray}}

\newcommand{\um}{\frac12}

\newcommand{\NP}[1]{Nucl.\ Phys.\ {\bf #1}}
\newcommand{\PL}[1]{Phys.\ Lett.\ {\bf #1}}

\newcommand{\PR}[1]{Phys.\ Rev.\ {\bf #1}}

\begin{document}
\begin{titlepage}
\vskip0.5cm
\begin{flushright}
DFTT 23/96\\
HUB-EP-96/24\\
\end{flushright}
\vskip0.5cm
\begin{center}
{\Large\bf The Spectrum of the 2+1 Dimensional}
\vskip0.2cm
{\Large\bf Gauge Ising Model}
\end{center}
\vskip 1.3cm
\centerline{V. Agostini$^a$, G. Carlino$^a$, 
M. Caselle$^a$\footnote{e--mail: caselle~@to.infn.it}
and M. Hasenbusch$^b$\footnote{e--mail: hasenbus@birke.physik.hu-berlin.de}}
\vskip 1.0cm
\centerline{\sl  $^a$ Dipartimento di Fisica 
Teorica dell'Universit\`a di Torino}
\centerline{\sl Istituto Nazionale di Fisica Nucleare, Sezione di Torino}
\centerline{\sl via P.Giuria 1, I-10125 Torino, Italy}
\vskip .4 cm
\centerline{\sl $^b$ Humboldt Universit\"at zu Berlin, Institut f\"ur Physik} 
\centerline{\sl Invalidenstr. 110, D-10099 Berlin, Germany} 
\vskip 1.cm

\begin{abstract}
We present a high precision Monte Carlo study of the  spectrum of the	
$Z_2$ gauge theory in $2+1$ dimensions in the strong coupling phase. 
Using state of the art Monte Carlo techniques we are able to accurately 
determine up to three masses in a single channel. We compare our results
with the strong coupling expansion for the lightest mass and with results for
the universal ratio $\sigma/m^2$ determined for the $\phi^4$-theory. 
Finally the  whole spectrum is compared with that obtained from the 
Isgur-Paton flux tube model and the spectrum of the $2+1$ dimensional
$SU(2)$ gauge theory. A remarkable agreement between the Ising and SU(2)
spectra (except for the lowest mass state) is found. 
\vskip0.2cm
\end{abstract}
\end{titlepage}

\setcounter{footnote}{0}
\def\thefootnote{\arabic{footnote}}

\section{Introduction}
This paper is devoted to the study of the glueball spectrum in the 
three dimensional gauge Ising model. There are two main reasons for which 
this model is interesting. The first one 
is that, as it is well known,  the infrared regime of
Lattice Gauge Theories (LGT) in the confining phase  displays a large 
degree of universality. Hence it is quite convenient to study such
universal infrared behaviour  in the case of the three dimensional 
Ising gauge model, which is the simplest 
(but nevertheless non-trivial) possible lattice gauge theory and allows high
precision Montecarlo simulations with a relatively small amount of CPU time.
The main evidence  in favour of  the above mentioned universality 
is given by the Wilson loop behaviour, whose functional form 
 does not depend on the choice of the gauge group, and shows a rather 
simple dependence on the number of space-time 
dimensions. Both these features are commonly understood as 
consequences of the fact that the relevant degrees of freedom in the  
confining regime are string-like excitations. Such a string description 
is still not well understood, the theory being anomalous at the quantum 
level. For instance, it is absolutely not clear how to translate at 
the string level the choice of the gauge group and which is the correct string
Lagrangian. However it is widely 
believed that in the infrared  regime, as the interquark distance increases 
all these different string theories 
flow toward a common fixed point which is not anomalous and corresponds
to the two dimensional conformal field theory 
of $(d-2)$ free bosons ($d-2$ being the number of transverse dimensions 
of the original string theory)~\cite{olesen}. 
The phenomenological models which try to keep into account this string-like
picture are usually known as ``flux-tube'' models. They have proved to be very
efficient in describing the behaviour of the Wilson loop and in particular
of its quantum fluctuations~\cite{lsw,lus81}. 
Moreover, the interquark potential that one obtains
with these models gives a  good description of the spectrum of
heavy quarkonia~\cite{ip,bbbpz}. As an example of the strategy discussed 
above  in~\cite{wloop,cgmv} some flux tube predictions on the behaviour of 
Wilson loops  were succesfully tested with high precision simulations 
in the three dimensional gauge Ising models. These studies allowed to obtain 
a clear insight in the fine structure of the flux tube model.

In the case of the glueball spectrum the situation is less clear. 
A string-inspired model exists also for the glueball 
spectrum: the Isgur Paton model (IP in the following).
However  in this case the equivalent of the interquark distance for the
Wilson loop is missing. In other words, it is not obvious to find a 
length scale  above which one can think to have reached 
a common universal infrared behaviour.

For this reason it would be very interesting to test if the same universality
(which, as mentioned above, should manifest itself as a substantial independence
from the choice of the gauge group) displayed by the Wilson loops also 
holds for the glueball spectrum. This can be more easily studied in the (2+1)
dimensional case, for which some relevant simplifications occur in the spectrum
(see below) and a much higher precision can be achieved in the Montecarlo
simulations. 
Recently the glueball spectrum was obtained for the (2+1) dimensional
SU(2) gauge model~\cite{tep1,tepmor}. In this
paper we shall present the spectrum in the case of the (2+1) Ising gauge
model. The comparison between the SU(2) and Ising spectra shows that, 
not only the pattern of the states is the same in the two models, but also 
the values of the masses (except for the lowest state) are in remarkable
agreement.
This is a strong evidence in favour of the above mentioned universality, and
suggests that the higher states of the glueball spectrum of any LGT, 
(as it happens for the behaviour of large enough Wilson loops)
 can be predicted by some relatively simple flux-tube inspired model.
The simplest possible model of this type is the IP model. So we shall devote
the last part of this paper to a detailed comparison with its predictions.

\vskip 0.3cm
A second important reason of interest in the gauge Ising model is that it is 
related 
by duality to the ordinary three dimensional spin Ising model. As a consequence,
the glueball spectrum is mapped in the spectrum of massive excitations  
of the spin Ising
model. The knowledge of the spectrum can teach us a lot on the physics of the
model (this is the lesson that we have learnt from the S-matrix approach to two
dimensional field theories) and also allows a  non trivial check of our
results with the predictions of the $\phi^4$ description of the spin 
Ising model. In particular we shall use 
the value of the lowest mass excitation: $1/\xi$ (which coincides with the
$0^+$ glueball state of the dual gauge model), which we can measure with 
very high precision, to test our simulations. We shall compare our results for 
the correlation length $\xi$ with
those obtained with the longest available strong coupling series and with some
recent calculations with the $\phi^4$ model. Indeed, the behaviour of the
correlation length in the broken phase of the 3d Ising model and its comparison
with the correlation length in the high temperature phase and with its second
momentum approximation is a very interesting problem in itself. We 
shall discuss this topic in detail in a separate paper~\cite{next}.

This paper is organized as follows:
Sect. 2 and 3 are devoted to a description of the model that we have studied and
of the details of our simulations. In sect. 4 we report our results on the lowest
mass excitation and compare them with some existing predictions. Sect. 5
is devoted to a description of our estimates for the full spectrum and to
a comparison with existing data on SU(2) and with the Isgur Paton model. 
The last section is devoted to some concluding remark.

\section{General Setting}
\subsection{The Models}

We take a lattice of $N_t$ ($N_s$) spacings in the time (space)
direction. Periodic boundary conditions are imposed in time and in space
directions. 
The gauge fields are described by the link
variables $g_{n;\mu} \in {-1,1}$, where $ n \equiv (\vec x,t)$
denotes the space-time position of the link and $\mu$ its direction.
We choose for simplicity the same bare coupling $\beta$ in the time and 
space directions. The Wilson action is then
\eq
S_{gauge} =  - \beta \;\; \sum_{n,\mu<\nu} \;\; g_{n;\mu\nu}
\label{Sgauge}
\en
where 
$g_{n;\mu\nu}$ are the 
plaquette variables, defined as usual by
\eq
g_{n;\mu\nu}=g_{n;\mu} \; g_{n + \mu;\nu} \;
g_{n + \nu;\mu} \; g_{n;\nu}~~~.
\label{plaq}
\en

The Ising spin model in 2+1 dimensions is defined 
by the action	
\eq
 S_{spin} = - \tilde\beta \sum_{n,\mu} s_n s_{n+\mu} \; , 
\label{Sspin}
\en
where the field variable $s_n$ takes the values $-1$ and $+1$. 

The Kramers--Wannier duality transformation relates the two partition 
functions:
\eqa
Z_{gauge}(\beta)~\propto~ Z_{spin}(\tilde\beta)&& \nonumber \\
{\tilde\beta}=-\um\log\left[\tanh(\beta)\right]~~&&~~. 
\label{dual}
\ena
Using the duality transformation it is possible to build 
up a one--to--one mapping of physical observables of the gauge system 
onto the corresponding spin quantities.

In the following we shall use  this duality transformation to relate the lowest
state of the spectrum of the gauge Ising model: the $0^+$ glueball, 
to the correlation length of the spin Ising model, for which several
predictions, both from strong coupling expansions and from the $\phi^4$ theory
exist. 

Another important consequence of duality is that the string tension of the
gauge Ising model is mapped into the interface tension of the Ising spin model.
In the following we shall denote both them with the same symbol $\sigma$.

\subsection{The algorithm}
For the  determination of the  spectrum in the broken 
phase of the gauge model we used a (gauge version) of the microcanonical demon
algorithm~\cite{creutz}. 

In our simulation the microcanonical demon algorithm of \cite{creutz} was 
combined with a particularly efficient canonical update 
of the demons \cite{Kari} in order to obtain the canonical ensemble of the 
gauge model.
This algorithm was implemented in the multispin coding technique.
The demons are auxiliary degrees of freedom that are added to the system. 
One considers the combined action 
\begin{equation}
S = \sum_{\alpha=1}^{bit} S_{gauge}^{\alpha} + S_D^{\alpha} \;\;, 
\end{equation}
where $bit=64$ independent copies of the gauge model and the demon 
system are simulated. 
The demon action is given by
\begin{equation}
S_D = \beta \sum_{n,\mu} d_{n;\mu} \;\; ,
\end{equation}
where the demon takes the values $0,2,..., d_{max}$. For our simulations
we have chosen $d_{max}=6$. The updates performed are similar to standard 
Metropolis updates.             
First a change of the sign of a given link variable $g_{n,\mu}$ is proposed. 
Then the corresponding change of the action of the gauge model 
$\Delta S_{gauge}  =S_{gauge}'-S_{gauge}$ is computed.
If $d_{n,\mu} - \Delta S_{gauge}$  is in the allowed range of the demon then
the proposal for the link variable is accepted and the value of the 
demon is changed to $d_{n,\mu}'= d_{n,\mu} - \Delta S_{gauge}$. 
Performing such updates once for all $\mu$ and $n$ is called one "sweep" 
in the following.   

In order to simulate the canonical ensemble of the model the microcanonical
updates are interleaved with canonical updates of the demon system. 
The demon with $d_{max}=6$ can be thought of as a sum of two two-state demons
with the energies $0;2$ and $0;4$ respectively. 

The demon-update is done separately for these two demon-species. 
First a new total number $E$ of exited demons is chosen with a probability 
proportional to the Boltzmann-weight 
\begin{equation}
 P(E) \propto  \frac{N_D!}{(N_D-E)! \; E!} \; \exp(- \bar{\beta} E)  \;\; , 
\end{equation}
where $N_D=64 \times  3 \times N_t \times N_s^2$ is the total number
 of demons,  $E$ the number of demons in the 
exited state and $ \bar{\beta} = 2 \beta $  or $ \bar{\beta} = 4 \beta $ for
the two demon species respectively. 

Next if $N_c = E_{new}-E_{old} > 0$ then $N_c$ randomly chosen demons with 
$d=0$ are updated  to $d=2,4$ respectively  else 
$-N_c$ randomly chosen demons with $d=2,4$ are updated to $d=0$. 

In addition we performed translations of the demons on the lattice and 
shifts along the 64 gauge systems. 

The  method
discussed above is, up to our knowledge, the first
attempt to construct a microcanonical demon algorithm for a gauge model
and  has proved to be a very powerful tool.
For moderate correlation lengths, as discussed in this paper, it
 should  provide superior performances compared to a 
cluster-update \cite{bkkls}. 
A careful quantitative study of this question is in progress \cite{ch}.

\section{The simulation}

\subsection{Determination of the Spectrum.}
In general the mass-spectrum of a theory is given by the eigenvalues of the 
Hamiltonian of the theory. The discrete version of the Hamiltonian on 
the lattice
is called transfer matrix $T$. For a finite lattice the transfer matrix of the 
Ising model is a real symmetric matrix. Therefore it can be 
diagonalized. Let us denote the resulting eigenvalues  by $\lambda_{i}$.
Then the mass-spectrum is given by 
\begin{equation}
m_i = -\log\left(\frac{\lambda_{i}}{\lambda_{0}}\right) \;\; ,
\end{equation}
where $\lambda_{0}$ is the largest eigenvalue of $T$.
For sufficiently small system sizes the eigenvalues of $T$
can be computed by numerically diagonalizing $T$. 
In the case of the 2+1 dimensional  Ising spin model one can reach at most 
lattices of size $5^2 \times \infty$
\cite{xxx} this way. 
For larger lattice sizes one has to rely on Monte Carlo techniques. 
The basic strategy is to compute expectation values of certain correlation
functions. Masses can then be determined from the decay of these 
correlation functions with the separation in time. 
\eqa
 G(t) &=&  \langle A(0) B(t) \rangle =  \frac{\langle 0| A T^t B |0 \rangle}
 {\langle 0| T^t |0 \rangle} \nonumber \\
&\sim & \frac{1}{\lambda_0^t}
\sum_i \;\; \langle 0| A |i\rangle 
\langle i| T^t |j \rangle \langle j| B |0 \rangle \;\; 
= \sum_i \;\;  c_i \left(\frac{\lambda_i}{\lambda_0}\right)^t \;=\;
 \sum_i c_i \exp(-m_i t)~~,
\label{expdec}
\ena
where $|i\rangle $ denotes the eigenstates of the transfer matrix and
\begin{equation}
 c_i =  \langle 0| A |i\rangle  \langle i| B |0 \rangle \;\;.
\end{equation}

The main problem in the numerical determination of masses  is to find
operators $A$ and $B$ that have a good overlap with a single state $|i\rangle$; i.e.
that $c_i$ is large compared with $c_j$, $j \ne i$.
The first, important, step in this direction 
is to realize that,
by choosing the symmetry properties of the operators $A$ and $B$ properly, we can
select channels; i.e. we can choose, for example, $A$ and $B$ such
that all the $c_i$'s vanish except those corresponding to, say,
a given value of the angular momentum.
Due to its importance, we devote the next section to a
detailed discussion of the choice and construction of the lattice operators.
By using  so called ``zero momentum''  operators,
namely operators obtained by summing over a slice orthogonal to the
time direction, all $c_i$'s that correspond to nonvanishing momentum 
vanish. 

A systematic way to further improve the overlap  
is to study
simultaneously the correlators among several
operators $A_{\alpha}$ that
belong to the same channel. 
This is indeed a natural 
prescription in the context of the glueball physics since the glueballs are
expected to be extended objects and choosing several extended operators on the 
lattice one can hope to find a better overlap with the (unknown) glueball wave
function. 
One must then measure all the possible correlations among these operators 
and construct the crosscorrelation matrix defined as:
\eq
\label{cross}
C_{\alpha \beta}(t)=\langle A_{\alpha}(t)A_{\beta}(0) \rangle - 
\langle A_{\alpha}(t)\rangle
\langle A_{\beta}(0) \rangle
\en

By diagonalizing the crosscorrelation matrix one can then obtain the mass
spectrum.

This method can be further improved\cite{kro,lw} by studying  the generalized 
eigenvalue problem
\eq
\label{lwge}
C(t)\psi=\lambda(t,t_0)C(t_0)\psi
\en
where $t_0$ is small and fixed (say, $t_0=0$). Then it can be shown
that the various masses $m_i$ are related to the generalized eigenvalues as
follows~\cite{kro,lw}:

\eq
\label{lwmf}
m_i=\log{\left(\frac{\lambda_i(t,t_0)}{\lambda_i(t+1,t_0)}\right)}
\en

where both $t$ and $t_0$ should be chosen as large as possible, 
$t>>t_0$ and as $t$ is varied the value of $m_i$ must be
 stable within the errors.
Practically one is forced to keep $t_0=0,1$ to avoid too large statistical
fluctuations and at the same time $t$ is in general forced to stay in the 
range $t=1$ to $7$, depending on $\beta$ and the channel, 
to avoid a too small signal to noise ratio.
This method is clearly discussed
in~\cite{kro,lw} and we refer to them for further details. All the results that
we shall list below for the glueball spectrum have been obtained with this
improved method.  
In order to give some informations on the reliability of the estimates
 we shall also list, besides the numerical values of the masses, the pair 
$(t_0,t)$  used to extract them.

\subsection{Choice of operators.}
The various glueball masses are labelled by their angular momentum.
Thus, in order to distinguish the various states of the spectrum one must 
construct operators with well defined angular momentum.  The
fact that we are working in (2+1) dimensions makes this construction rather non
trivial, and definitely different from 
the (3+1) dimensional case.
Moreover, since we are working on a cubic lattice, 
where only rotations of multiples of 
$\pi/2$ are allowed, we must study the symmetry
properties of our operators with respect to a finite subgroup of the
two-dimensional rotations. 
Let us first ignore the effect of the lattice
discreteness and deal with the peculiar features which, already in the
continuum formulation,  the (2+1) Ising (and SU(2), which is completely
equivalent in this respect) spectrum has with respect to the (3+1) dimensional
SU(3) spectrum.
These features have been discussed  for the SU(2) model in 
ref.~\cite{tep1,tepmor} 
to which we refer for further details.
We shall summarize here only the main results.
\begin{description}
\item{a]} For the Ising model, as for the SU(2) model, we cannot define a charge
conjugation operator. The glueball states are thus labelled only by their 
angular momentum $J$ and by their parity eigenvalue $P=\pm$. 
The standard notation is $J^{P}$

\item{b]} In (2+1) dimensions it can be shown that all the states with 
angular momentum different from zero are degenerate in parity. Namely $J^+$ and
$J^-$ (with $J\neq 0$) must have the same mass.

\end{description}

On the cubic lattice the group of two dimensional rotations and reflections 
becomes the dihedral group $D_4$. 
This group is non abelian, has eight elements and five
irreducible representations. Four of these are one-dimensional irreps, the last
one has dimension two. The group structure is completely described by the
table of characters~\cite{ham} which we have reported in tab.1 .

 \begin{table}[ht]
 \caption{ \sl Character table for the group $D_4$}
 \label{character}
  \begin{center}
   \begin{tabular}{|l|c|r|r|r|r|}
   \hline
  & ${\bf 1}$ & $C_4^2$ & $C_4(2)$ & $C_2(2)$ & $C_{2'}(2)$ \\
   \hline
 $A_1$ & $1$ & $1$ & $1$ & $1$ & $1$ \\
 $A_2$ & $1$ & $1$ & $1$ & $-1$ & $-1$ \\
 $B_1$ & $1$ & $1$ & $-1$ & $1$ & $-1$ \\
 $B_2$ & $1$ & $1$ & $-1$ & $-1$ & $1$ \\
 $E$ & $2$ & $-2$ & $0$ & $0$ & $0$ \\
   \hline
   \end{tabular}
  \end{center}
 \end{table}

In the top row of tab.1 are listed  the invariant classes of the group, and 
in the first column the irreducible representations. We have followed the
notations of~\cite{ham} to label classes and representation (with the exception
of the class containing the identity which we have denoted with ${\bf 1}$
instead of the usual $E$ to
avoid confusion with the two-dimensional representation). The entries of the
table allow to explicitly construct the various representations and hence also 
the lattice operators 
which we are looking for.
The relationship of these operators with the various glueball 
states immediately follows from the group structure. In particular one can show
that:

\begin{description}
\item{a]} Only operators with angular momentum $J~(mod(4))$ can be 
constructed.
This is a common feature of all cubic lattice regularizations. It means that 
glueball states which in the continuum have values of $J$ higher than 3
appear on the lattice as secondary states in the family of the 
corresponding $J~(mod(4))$ lattice operator.

\item{b]} 
The four one-dimensional irreps are in  correspondence with the
$J$ even states. More precisely:
$$  0^+ \to A_1,\hskip 1cm 0^- \to A_1,\hskip 1cm 
 2^+ \to B_1 ,\hskip 1cm
 2^- \to B_2$$

This means that the discreteness of the lattice splits the degeneracy between
$2^+$ and $2^-$ which we discussed above. The splitting between these two states
gives us a rough estimates of relevance of the breaking of the full rotational
group due to the lattice discretization. As we shall see below this splitting is
essentially zero within the errors, 
in agreement with our expectation that approaching
the continuum limit the full continuum symmetries should be recovered.
Notice however that this is a very non-trivial result since the operators
associated to $2^+$ and $2^-$ on the lattice turn out to be very different.

\item{c]}
All the odd parity states are grouped together in the two-dimensional
irreducible representation $E$. This means that we cannot distinguish among 
them on the basis of the lattice symmetries. We can conventionally assume, say,
that the $J=3$ states have a mass lower than the $J=1$ ones (this would agree
with the pattern which emerges from the Isgur-Paton model, see below), and that 
the $J=1$ thus appear as secondaries in the $J=3$ family. However in the
following we shall avoid this assumption and shall denote the states belonging
to this family as $J=1/3$ states. In agreement with the above discussion, if the
full rotational symmetry is recovered, we expect that the states belonging to
this family are degenerate in parity and thus that the lowest mass states, which
are the ones that we can measure more precisely appear as a doublet. This
prediction will be in fact confirmed by our results.

\end{description}

The simplest lattice operators, constructed according to the character table,
are shown in fig.1. 
 Notice however that these are only a small subset of the
operators that we actually measured in our simulations. In particular we
studied 27 different operators for the $0^+$ channel, and 15, 9, 5 ,16 for the
$2^+,2^-,0^-$ and $1/3$ channels respectively.

\subsection{The data sample.}

First we carefully checked for $\beta = 0.72484$  the dependence of the
masses on the lattice size. In table \ref{finit} we give results for the 
$0^+$ channel for various spatial extensions $N_s$ of the lattice. The results
become stable within our numerical accuracy for $N_s \ge 16$. 

 \begin{table}[ht]
 \caption{\sl Study of finite size effects for the $Z_2$ gauge
  model at $\beta=0.72484$ (which corresponds to $\beta=0.2391$
  for the spin model). 10000 measurements, $t=3$. $N_t = 48$ throughout.
            }
 \label{finit}
  \begin{center}
   \begin{tabular}{|c|c|c|l|l|}
   \hline
 $N_s$ &$\xi_{0+}$&$\xi_{0+}'$&$\xi_{0+}''$ \\
   \hline
  8 &  1.485(3)&  0.699(7) &  0.59(5)    \\
 10 &  1.408(3)&  0.836(6) &  0.52(2)    \\
 12 &  1.313(3)&  0.800(7) &  0.58(2)    \\
 16 &  1.298(3)&  0.721(8) &  0.53(3)    \\
 20 &  1.295(3)&  0.717(8) &  0.52(3)    \\
 24 &  1.296(3)&  0.705(8) &  0.52(3)    \\
   \hline
   \end{tabular}
  \end{center}
 \end{table}

In order to be 
safe of finite size effects we have chosen $N_s \approx 15 \xi_{0+}$ for 
the following simulations. We denote with $N_t$ the ``time'' direction in 
which we calculate the correlations. In these simulations we have always 
chosen  $N_t>N_s$.

Next we performed simulations for $\beta=0.74057$, $0.74883$, $0.752023$ and 
$0.75632$ (the corresponding (dual) couplings for the spin model are reported in
tab.~\ref{data1})
 on lattices of the size $72 \times 36^2$ , $60\times 40^2$, 
$70\times 50^2$ and
$100 \times 70^2$.  
We performed 4790, 11890, 5000 and 2080
measurements on each of the 64 systems for the four $\beta$
values respectively.
A measurement was performed after 200, 200, 400 and 500 update-sweeps.
We made no attempt to study carefully the autocorrelation times of
Wilson-loops.  Some rough tests indicate  that they are 
of the same order as the number of update-sweeps
which we performed for one measurement. 
All the information on our simulations are summarized in tab.~\ref{data1}.

The total amount of CPU-time for the run of the $100 \times 70^2$ lattice 
took about $55$ days on a DEC 3000 Model 400  AXP Workstation. 
The ratio of CPU time spent in updating and measurement was about $3/1$.

 \begin{table}[ht]
 \caption{\sl Some informations on the data sample for the glueball spectrum:
$\beta$ is the coupling constant of the gauge Ising model and $\tilde\beta$ the
corresponding (dual) coupling for the spin model. $N_t$ and $N_s$
 are the lattice
sizes. With u.s. we denote the number of update sweeps between two measures and
with m.s. the total number of measured sweeps.}
 \label{data1}
  \begin{center}
   \begin{tabular}{|c|c|c|c|c|}
   \hline
 $\beta$ & $\tilde\beta$ & $N_t\times N_s^2$ & u.s. & m.s. \\
   \hline
 0.74057 &  0.23142& $ 72\times 36^2$ &  200 &  4790    \\
 0.74883 &  0.22750& $ 60\times 40^2$ &  200 & 11890    \\
 0.75202 &  0.22600& $ 70\times 50^2$ &  400 & 5000    \\
 0.75632 &  0.22400& $100\times 70^2$ &  500 & 2080    \\
   \hline
   \end{tabular}
  \end{center}
 \end{table}

Our results are collected in  table 4.
The data presented correspond to the inverse masses: 
$\xi_i\equiv\frac{1}{m_i}$. In the first row we have reported the dominant state
for each family while in the subsequent rows are reported the various secondary
excitations. For each mass we have also reported the pair  $(t_0,t)$
used to extract the mass from the data. When the statistical fluctuations were
too large to obtain a meaningful statistical error, we tried to fix anyway an 
approximate estimate for the masses. In the tables these estimates are reported
in square brackets, to remind that they are certainly affected by large errors.
Empty boxes denote the cases in which even this approximate estimate was 
impossible to find. Looking at the data one can immediately see the announced
degeneration of the $2^+$ and $2^-$ 
families and the fact that the first two states of
the 1/3 family appear as a doublet.

\begin{table}[ht]
\label{gbs1}
\caption{\sl Glueball spectrum. The discussion is given in the text.}
  \begin{center}
\begin{tabular}{|c|c|c|c|c|c|}
\hline
$\beta$ &  $ 0^+$     & $0^-$   & $2^{+}$   & $2^{-}$ & $1/3$ \\
\hline
0.74057 &  (0,3)     & (0,2)   &  (0,2)   & (0,2)   & (0,1)       \\
        & 1.864(5)   & 0.60(5) &  0.77(1) & 0.75(1) & [0.46]  \\
 \hline
        & (0,3)     & (0,1)   &  (0,2)    & (0,2)  & (0,1)   \\
        &1.03(2)    & $$ [0.42] &  0.65(2) & $$ [0.60]& $$ [0.46] \\
 \hline
        & (0,2)     & (0,1)  &   (0,2)    &   (0,1)  & (0,1)      \\
        &0.77(2)    & $$ [0.37] & $$ [0.60] & $$ [0.44] & $$ [0.38] \\
 \hline
 \hline
0.74883 & (0,3)(1,2) &(0,2)(1,2)&(0,2)(1,2) & (0,3)(1,2)& (0,2)(1,1)\\
        & 2.592(5)   & $$ [0.75] &  1.04(2)  & 1.05(3)   & 0.66(3)\\
  \hline
        &  (0,2)(0,3) &         & (0,2)      &  (0,2)(1,2)& (0,2)(1,1) \\
        &  1.42(1)    &         & $$ [0.82]   &  0.82(3)  &  0.63(3)  \\
   \hline
        &  (0,3)      &         &            & (0,2)(1,2)& (0,1)(1,1) \\
        & 1.06(2)    &         &            & $$ [0.63]  & $$ [0.50] \\
    \hline
     \hline
0.75202 & (0,4)(1,3) &  (0,3)  & (0,3)(1,3)&  (0,3)(1,3)& (0,3)  \\
       &3.135(9)   &  0.96(7)& 1.27(4)   &  1.23(3)   & 0.8(1) \\
 \hline
       & (0,4)(1,2) & (0,2)  & (0,3)(1,3) & (0,4)(1,3) & (0,3)  \\
       &1.72(1)   &0.69(3) & 1.00(5)    &   1.0(1)   & 0.8(1) \\
\hline
      &	 (0,4) &      (0,2)   & (0,1)(0,2) & (0,2)     & (0,2)  \\
      &	1.20(4)    & $$ [0.65] & $$ [0.8]  & $$ [0.72]   & $$ [0.62] \\
\hline
\hline
0.75632 & (0,7)(1,4)& (0,2)   & (0,4)(1,2)& (0,4)    & (0,3)(1,2)\\
        & 4.64(3)  & $$ [1.1] &  1.76(3)  & 1.69(3) &   1.2(1)  \\
\hline
        & (0,5)(1,3)& (0,1)   & (0,3)(1,2)&(0,3)(1,2)&  (0,3)(1,2) \\
        & 2.50(3)  & $$ [0.80]&  1.39(3)  & 1.38(5)&  1.1(1) \\
\hline
        & (0,4)(1,2) & (0,1)   &  (0,3)   & (0,2)(1,2)&  (0,2)(1,1)\\
        & 1.72(4)   & $$ [0.58]& $$ [1.07] & 1.1(1)     &  $$ [0.9]  \\
\hline
\end{tabular}
  \end{center}
\end{table}

\section{The lowest mass state}
\subsection{Comparison with strong coupling expansions}

In this section we concentrate on the lowest mass state: the $0^+$ glueball.
As mentioned above this state  coincides with the inverse of the correlation
length $\xi$ of the spin Ising model. For this observable, very long strong
coupling series have recently been obtained by
 H.Arisue and  K.Tabata~\cite{at}. To extract from these series estimates for
 $\xi$ at the values of $\beta$ in which we are interested in
 we  used the so called ``double biased
inhomogeneous differential approximants'' (IDA). The techniques of IDA
 is well described in~\cite{ida,lf}, to which
we refer for notations and further details. In order to keep the fluctuations 
of the results under control, we have chosen to use double biased IDA, namely 
we have fixed both the
critical coupling and $\nu$ to their most probable values $\beta_c=0.221655$
and $\nu=0.63$~\cite{nu} 
(for a discussion  of these double biased IDA see again
ref.~\cite{lf}). Following ref.~\cite{lf} we use 
the notation [K/L;M] for the approximants.  In 
fig.2
approximants of $\xi$ at $\beta=0.75632$ are given for a 
large range of [K/L;M]. 
Certainly the values for $\xi$ are centered around a value of about $4.6$, which
is consistent with our Montecarlo result. We made no attempt to extract an 
error-estimate from the spread of the values of the IDA's. 

In table 5 results for $\beta=0.72484$, $0.74057$, $0.74883$,
$0.75202$ and $0.75632$
for three choices of K,L,M 
which give estimates consistent with our Montecarlo estimate for 
$\beta=0.75632$ are summarized.

For comparison we give in addition the values of $\xi$ in the last column,
which are taken from tab.~\ref{finit} and table 4.
\begin{table}[ht]
\label{data5}
\caption{\sl Comparison with the strong coupling expansion}
  \begin{center}
\begin{tabular}{|l|c|l|l|l|l|}
\hline
 $ \beta$ & $\tilde\beta$  &   [4/6;7] &  [6/6;5] &  [6/5;6]  &   $ \xi$ \\
\hline
0.72484  & 0.23910  &  1.2995  &  1.2998  &  1.2977   &  1.296(3) \\
0.74057  & 0.23142  &  1.8739  &  1.8751  &  1.8699   &  1.864(5) \\
0.74883  & 0.22750  &  2.5946  &  2.5974  &  2.5876   &  2.592(5) \\
0.75202  & 0.22600  &  3.1299  &  3.1343  &  3.1208   &  3.135(9) \\
0.75632  & 0.22400  &  4.6237  &  4.6325  &  4.6084   &  4.64(3)   \\
\hline
\end{tabular}
  \end{center}
\end{table}
The important observation is, that these three approximants also give results 
that are consistent with our Montecarlo estimates for
$\beta=0.72484$, $0.74057$, $0.74883$ and $0.75202$.

\subsection{Scaling behaviour and $\phi^4$ theory}

In the neighbourhood of 
the critical temperature the correlation length in the low temperature phase
of the spin Ising model scales as 
\begin{equation}
\label{scaling}
\xi \sim  f_{-} \left(\frac{\tilde\beta-\tilde\beta_c}
{\tilde\beta_c} \right)^{-\nu}~~~~.
\end{equation}
  Most of the recent 
estimates for the critical exponent $\nu$ range from $0.629$ to $0.632$ 
\cite{nu}, however in a recent MCRG study $\nu=0.625(1)$ was found 
\cite{gupta}. 
This
uncertainty can be eliminated by looking at adimensional combinations like
 $\sigma \xi^2$ (where $\sigma$ denotes the interfacial
tension in the spin model or, equivalently, the string tension in the gauge
model). In the scaling limit this product should be constant.
An interesting feature of this adimensional product is that
its value in the continuum limit can be evaluated in the framework of the
$\phi^4$ theory~\cite{m90}, which gives:

\eq
\sigma\xi^2=0.1024(88)
\label{phi4}
\en
Recently, very precise estimates of $\sigma$ have been obtained in the
same range of $\beta$ values in which we are 
interested~\cite{hp,cfghpv}\footnote{The values of $\sigma$ for $\tilde\beta=0.
2391, 0.2314$ and $0.2260$ are obtained by interpolation of nearby values. The
quoted errors take also into account the systematic uncertainty due to the fact
that we only know the first non-Gaussian contributions in the functional
form of the interface tension (see~\cite{cfghpv} for a discussion).} .
The results obtained by using these values and our estimates of $\xi$ 
are reported in tab.6. 

\begin{table}[ht]
\label{ratiosig}
\caption{\sl Scaling behaviour of $\xi$. In the second column we report 
the values of $\sigma$ extracted from [26] and [27].}
  \begin{center}
\begin{tabular}{|l|l|l|}
\hline
 $\tilde\beta$  & $\sigma$ &  $\sigma\xi^2$    \\
0.23910  &  0.05558(10)  &   0.0934(6) \\
0.23142  &  0.02752(10)  &   0.0956(9) \\
0.22750  &  0.01473(10)  &   0.0990(11) \\
0.22600  &  0.01011(10)  &   0.0994(16) \\
0.22400  &  0.00478(10)  &   0.1029(34)   \\
\hline
\end{tabular}
  \end{center}
\end{table}
The value of $\sigma\xi^2$ increases as $\beta$ approaches its critical
value.
This trend is the signature of the presence of a correction to 
scaling term. This type of corrections is always present (see for instance the
discussion in~\cite{lf}) but becomes visible only if the data sample is precise
enough.
We expect the following behaviour for $\sigma\xi^2$ :
\eq
\sigma\xi^2=(\sigma\xi^2)_0~ +~ \rho \xi^{-\omega}
\label{ip4a}
\en
where $\omega \approx 0.8$ \cite{nu}.
 $(\sigma\xi^2)_0$ is the continuum limit value  and $\rho$ an unknown
 constant.
Fitting our five data with eq.(\ref{ip4a}) we find:
$\rho=-0.015(3)$ and $(\sigma\xi^2)_0=0.1056(19)$ with a  reduced
$\chi^2$:  $\chi^2_{red}=0.42$ which corresponds to a confidence
level of $70\%$. This continuum limit value for $(\sigma\xi^2)_0$ 
is in perfect agreement with the $\phi^4$ prediction of eq.(\ref{phi4}). 
It also implies that
$m_{0^+}/\sqrt{\sigma}=3.08(3)$. We shall use this result in the following
section. Recently S.-Y. Zinn and M.E. Fisher~\cite{zf} 
published a rather smaller 
result: $\sigma\xi^2=0.0965(20)$, based on the data of ref.~\cite{hp} and 
a low temperature expansion. 
However most of the discrepancy between their result and
our one can be explained by the fact that they use, in contrast to us, 
the second moment correlation length.

\section{The whole glueball spectrum}
\subsection{Numerical Results}
In order to obtain a meaningful result for the continuum theory we have to 
measure the various masses of the spectrum in  physical units. 
Commonly the (square root of the) string tension is used for this
purpose.
Our results are reported in
tab.~7 where 
we used the $\sigma$-values listed in tab.6.
It can be seen that the scaling behaviour is well established
within the statistical errors for the whole spectrum 
(except for the lowest state that we
have discussed in the previous section). 
Thus, to extract the continuum values (except for $0^+$)
we have chosen the simplest possible attitude,
and took the weighted mean of the masses at the four values of $\beta$ as our
final result.
If we assume that corrections to scaling are of about the same size as for the
$0^+$ state, where  we have the by far most accurate results, this procedure
is justified.
These continuum values are reported in the
sixth column of tab. 7. In the last column we have reported, for the
reader's convenience the continuum values of the masses measured in units of the
lowest mass $0^+$. With the notation $(J^P)'$ and $(J^P)''$ we denote the second
and third excitation in the $J^P$ channel.
The continuum value for the $0^+$ state, reported in tab.7 is
that extracted in the previous section.

\begin{table}[ht]
\label{ip5a}
\caption{\sl The glueball spectrum
 in units of the square root of the string tension.
 In the sixth column
we report our extrapolation to the continuum limit. In the last column 
we report the same quantity, measured in units of 
the lowest excitation.}
  \begin{center}
\begin{tabular}{|l|l|l|l|l|l|l|}
\hline
 $J^P$ & $\beta=0.74057$ & $\beta=0.74883$ & $\beta=0.752023$ 
& $\beta=0.75632$  & $m/\sqrt{\sigma}$ &  $m/m_{0^+}$    \\
\hline
$0^+$  & 3.23(1)  & 3.18(2)  &  3.17(1)  & 3.12(5)  & 3.08(3) &  1 \\
$(0^+)'$  & 5.85(12)  & 5.80(6)  &  5.78(6)  & 5.79(13)  & 5.80(4)& 1.88(2)\\
$(0^+)''$  & 7.8(2)  & 7.8(2)  &  8.3(3)  & 8.4(3)  &  7.97(11)& 2.59(4) \\
$(0^-)$    & 10.0(7) &         &  9.9(8)  &         & 10.0(5)  & 3.25(16)\\
$(0^-)'$   &         &         &  13.8(6) &         & 13.8(6)  & 4.48(20) \\
$2^+$  & 7.8(1)  & 7.9(2)  &  7.8(3)  & 8.2(2)  &  7.88(8)& 2.56(4)\\
$2^-$  & 8.0(1)  & 7.8(2)  &  8.1(3)  & 8.6(2)  &  8.07(8)& 2.65(4)\\
$(2^+)'$  & 9.3(3)  &   &  9.9(5)  & 10.4(3)  &  9.86(20)& 3.23(7) \\
$(2^-)'$  &   & 10.0(4)  &  9.9(9)  & 10.5(5)  &  10.16(30) & 3.33(10)\\
$1/3$  &   & 12.5(6) &  12.4(1.6)  & 12.1(1.1)  &  12.4(5) & 4.07(16)\\
$(1/3)'$  &   & 13.1(6) &  12.4(1.6)  & 13.2(1.3)  &  13.0(5) & 4.26(16)\\
\hline
\end{tabular}
  \end{center}
\end{table}
As already discussed above, our data clearly confirm the degeneration between
states of different parity, for $J>0$. 

If we assume this degeneration as an
input information we can obtain more precise estimates for the spectrum, fitting
together all the $2^+$ and $2^-$ states and the first pair of $1/3$ states.
The resulting spectrum is reported in tab.9 .


Besides these states we also obtained estimates for the third excitation in the
$2^+,2^-,1/3$ and $0^-$ channels (see tab.4). However
these estimates are affected by a large uncertainty and we
did not try  to extrapolate a continuum limit value for the corresponding masses.

\subsection{The Isgur-Paton model}

More than ten years ago  Isgur and Paton suggested to describe the glueball
spectrum of a generic gauge theory within the framework of a flux tube model.
Up to our knowledge this is the only model which also incorporates the effects
of the quantum fluctuations of the flux tube in a consistent way (see below).
 We shall describe the IP model, directly in
the (2+1) dimensional case in which we are interested. For a more detailed
discussion and for the extension to (3+1) dimensions we refer to the original
paper~\cite{ip} and to~\cite{tepmor}. 
An appealing feature of the IP models is its simplicity. The starting point is
the assumption that the glueballs are ``rings of glue'', namely that they are
described by thin, oscillating, flux tubes which close upon themselves. This is
the natural extension to the glueball case of the flux tube model which
describes the mesons as flux tubes which end on quarks. 
The second, important, assumption of the model 
is the so called ``adiabatic approximation'' (which allows to 
explicitly quantise it). To understand this approximation, 
let us think for simplicity of a circular flux loop of radius $R$. It can have
two type of fluctuations: phonon-like fluctuations at a fixed radius, and
collective radial excitations. In the adiabatic approximation
these two types of excitations are treated in a completely different way. 
The phonons are quantised as free harmonic oscillators, neglecting the radial
excitations. Taking into account their contribution we can construct the total
energy $E_{tot}$ of the flux loop. Then we quantise the radial excitations using
the energy provided by the phonons as a fixed background potential. Thus the
problem of finding the mass eigenstates is reduced to the solution of a 
suitable radial Schr\"odinger equation in the only remaining
variable $R$. To this end let us first explicitly write $E_{tot}$.
The various phonons can only have frequencies which are multiples of $1/R$. 
Since the circular loop is embedded in the 2+1 dimensional space, a
phonon of frequency $m/R$ will imply an angular momentum for the glueball of 
$J=\pm m$ (the sign depending on the clockwise or counter-clockwise nature of the
phonon). Hence the total angular momentum due to the phonons is
\eq
J=\sum_m m(n^+_m-n_m^-)
\label{ipj}
\en
where $n^+_m,n_m^-$ is the number of phonons with angular momentum $+m$ and 
$-m$ respectively. 
The corresponding  energy is

\eq
M=\sum_m m(n^+_m+n_m^-)
\label{ipm}
\en

Some remarks are in order at this point.

\begin{description}
\item{a]}
The role of the parity operator in this context is to interchange + and $-$
phonons. Thus we see that (for $J> 0$)  excitations with the same angular
momentum but different parity  are degenerate in energy. 

\item{b]}
As it always happens when looking to the quantum fluctuations of a flux tube,
the zero point energy of the oscillators gives a divergent contribution 
to the total energy. This divergence can be regularized for instance with the
$\zeta$ function regularization. The resulting, finite, contribution to the
total energy is $13\pi/l$ where $l=2\pi R$ is the loop length. This term is the
equivalent in this context of the well known ``L\"uscher'' term which is 
present in the flux tube models for the interquark potential~\cite{lsw,lus81}.

\item{c]} In writing the total energy of the flux loop we must also take into
account, in analogy to the interquark case, a contribution proportional to the
length of the loop and to the string tension: $2\pi\sigma R$ and the 
possible existence of a constant term: $c_0$.

\item{d]} In the sums of eq.(\ref{ipj},\ref{ipm}) we must neglect the m=1
phonons, since the oscillations with $m=1$ are equivalent to infinitesimal 
rotations and translations. This has the important consequence that it is more
difficult to construct a $J=1$ or $J=3$ excitation than the $J=2$ one, which
hence has a lower mass. 
This is indeed confirmed by the simulations and is one of
the most appealing results of the IP model. The realization of the various
angular momenta, with the smallest possible total energy are listed in
tab. 8.

\begin{table}[ht]
\label{ip5e}
\caption{\sl
Phonon content of the various glueballs. The states are listed in order
of increasing energy. For some states we have listed some possible different
realizations.}
  \begin{center}
\begin{tabular}{|l|l|l|l|l|}
\hline
 $J^P$ & $ n_2$ &  $n_3$ &  $n_4$ & M   \\
\hline
$0^+$  &  0 & 0 & 0 & 0 \\
$2^{\pm}$  &  1 & 0 & 0 & 2 \\
$3^{\pm}$  &  0 & 1 & 0 & 3 \\
$0^+$  &  2 & 0 & 0 & 4 \\
$1^{\pm}$  &  1 & 1 & 0 & 5 \\
$0^+$  &  0 & 2 & 0 & 6 \\
$2^{\pm}$  &  1 & 0 & 1 & 6 \\
$0^-$  &  2 & 0 & 1 & 8 \\
\hline
\end{tabular}
  \end{center}
\end{table}

\end{description}

Taking into account all these features the total energy turns out to be:
\eq
E_{tot}(R)=2\pi\sigma R+ c_0 + \frac{M-13/12}{R}
\label{ip5etot}
\en

However in working out eq.(\ref{ip5etot}) we have considered the flux tube as an
``ideal'' string without thickness. We know that this is not the case. In fact
such  finite thickness of the flux tube has been recently observed with  
high precision exactly in the (2+1) dimensional Ising model that we are 
studying now~\cite{cgmv} and has been found to be (for the range of scales 
typical of these glueball states)
 of the order 
of $\frac{1}{\sqrt\sigma}$ (see below for a more detailed discussion). 
This is actually a general feature of gauge models and a
similar behaviour is expected also, for instance,
for (3+1) dimensional SU(3) gauge theory. This fact is taken into account in the
IP model, following the original suggestion in~\cite{ip}, by
introducing a phenomenological parameter $f$ which should soften the $1/R$
divergence as $R\to 0$. The final form of $E_{tot}$ is thus:

\eq
E_{tot}(R)=2\pi\sigma R+ c_0+ \frac{M-13/12}{R}(1-\exp[-f\sqrt{\sigma}R])
\label{ip5etot2}
\en

The resulting radial Schr\"odinger equation can then be solved numerically so as
to obtain the glueball spectrum and compare it with the Montecarlo results.

\subsection{Comparison with Ising and SU(2) spectra and discussion.}

The glueball spectrum that one obtains by solving the Schr\"odinger equation 
 depends on the two
phenomenological parameters $c_0$ and $f$, and in principle one could try to
tune these parameters so as to fit the results of the Montecarlo data. However
there are some features of the model which are largely independent from these
parameters and allow to discuss the reliability of the model in a more
general way.  We have reported in the fourth column
of tab.9  the IP spectrum obtained by choosing the
``ideal'' string model with $f=\infty$ and $c_0=0$
\footnote{ A suitable tuning of $f$ and $c_0$
could  lead to a better agreement with the Montecarlo results, 
which however would probably be fictious and  misleading.}.
This could help 
 to follow the forthcoming analysis and the comparison
between the IP model and our Montecarlo results.

\vskip 0.2cm
Let us first distinguish the splitting among different angular
momenta from that among different radial excitations at fixed $J$.
The pattern of the  states with different angular momenta 
can be already seen, without solving the Schr\"odinger equation, 
 looking at the 
$E_{tot}$ behaviour as a function of $J$. In fact the $E_{tot}$ pattern is 
essentially respected by the final mass spectrum. In this respect we can see
some peculiar features of the IP model (see tab.8):

\begin{description}
\item{1]}
The parity degeneration of the spectrum is present also in the IP model, and is
one of its constituent features.
\item{2]}
The lowest excitation (neglecting the radial excitations) 
above the $0^+$ is the $2^\pm$, while the $3^\pm$ doublet (which are the lowest
states in the $1/3$ family) has a higher mass. 
\item{3]} The $0^-$ state has a mass which is much larger than that of the
$3^\pm$ doublet.
\end{description}

While the first two features agree very well with the Montecarlo results (and in
particular the second point is probably the most appealing feature from a
theoretical point of view of the IP model), the third one completely disagrees.
Since it is a structural and unavoidable consequence of the IP model, it
indicates that some drastic improvement of the model is needed to reach a more
realistic description of the glueball pattern\footnote{Notice, as a side remark,
that the effect of the $f$ parameter is to increase 
the masses of the $0+$ family
and to decrease that of all the other angular momenta, as $f$ is decreased from
$\infty$ to 0. Thus point (3) above is not affected by $f$.}

\vskip 0.3cm
Let us now turn to  
  the splitting of the radial excitations. By solving the  
the Schr\"odinger equation  one can see the following features (see tab.9):
\begin{description}
\item{1]}
The radial splittings  are, in general, smaller than those between different
angular momentum families. This agrees with the adiabatic approximation, which
however should require the radial splitting to be much smaller than the 
angular
momentum one, while in the spectrum the difference is only of a factor of about
two.
\item{2]} The radial splittings show a small monotone dependence on the value of
the masses. They become smaller and smaller as the masses increase. Since the
overall values of the masses at a fixed $J$ changes as $f$ is changed, also the
radial splittings show a corresponding small dependence on $f$.
\item{3]} Comparing the splittings predicted by the IP model and those found in
the Montecarlo simulation we see that the order of magnitude is essentially the
same, the IP ones being in general a bit larger than the Montecarlo ones. 
\end{description}

\begin{table}[ht]
\label{ip5f}
\caption{\sl Comparison between the  Ising, SU(2) and IP spectra. In the last
column we report the IP predictions for the glueball radii.}
  \begin{center}
\begin{tabular}{|l|l|l|l|l|}
\hline
$J^P$ &  Ising & SU(2) & IP & $R_{max}\sqrt\sigma$    \\
\hline
$0^+$  &  3.08(3)& 4.763(31) & 2.00 & 0.50 \\
$(0^+)'$ & 5.80(4)&   & 5.94 & \\
$(0^+)''$  &  7.97(11)& & 8.35 &  \\
$2^{\pm}$  &  7.98(8) & 7.78(10)& 6.36 & 0.65 \\
$(2^\pm)'$  &  9.95(20)&  & 8.76 & \\
$0^-$   &   10.0(5) & 9.90(27) & 13.82 & 1.20\\
$(0^-)'$  & 13.8(6) &   & 15.05 & \\
$(1/3)^\pm$   & 12.7(5)& 10.75(50) & 8.04 & 0.75 \\
\hline
\end{tabular}
  \end{center}
\end{table}

Looking at tab.9 we see that the major disagreement in the mass values is for
the lowest state which is predicted to be too light. This is certainly a
consequence of the fact that in this ideal string picture we have introduced no
self repulsion term for the flux tube. Thus the lower states, and in particular
the $0^+$ ground state, are smaller (and
consequently lighter) in the IP model than in the actual gauge model. It would
be interesting, to test this picture, to have some idea of the 
``mean radius'' of the various glueball. This is indeed possible in the
framework of 
 the IP model, since from the numerical solution of
the radial Schr\"odinger equation one can extract not only the eigenstates but
also the eigenfunctions of the various glueball states.
They give us an idea of the actual 
sizes of the various glueballs which should coincide with the position
$R_{max}$ of the maxima in the corresponding probability distributions . 
The results for the lowest excitations (with
 the same choice $c_0=0$ and $f=\infty$ as above: ``ideal'' string), measured
 in units of $1/\sqrt\sigma$ are 
reported in the last column of tab.9 .
These eigenfunctions are plotted (with the radial coordinate measured in units 
of $1/\sqrt\sigma$) in fig.3 for the lowest excitation in each channel, while in
fig.4 we have plotted the first three excitations in the $0^+$ channel.

The fact that we have studied in particular the Ising model, allows us to give
a somehow more quantitative evidence in favour of the above discussion. In fact
a lower bound for the radius at which the free flux tube picture should break 
down, and some self-interaction of the flux tube must be taken into account, is
certainly given by the thickness of the flux tube itself.  The value of this
thickness has been studied in great detail, in the case of the three dimensional
Ising model in~\cite{cgmv}. It turns out that
the thickness $\rho$ is a function of the length of the tube (in our
case $2\pi R$).
and that the functional form of the dependence of  $\rho$
on $R$ is an universal function which only depends on the type of boundary
conditions of the flux tube. In the case in which we are interested the law 
is~\cite{lus81,cgmv}:
\eq
\sigma\rho^2=\frac{1}{2\pi} \log({2\pi\sqrt\sigma R}) +b
\label{rho}
\en
where  $b$ is a non-universal constant which depends on the details of the
model. In  the case of the Ising model this constant has been measured with
high precision in~\cite{cgmv}. Plugging its value in eq.(\ref{rho}) and
choosing $R\sqrt\sigma=0.50$, namely the radius predicted by the IP model for 
the $0^+$ glueball, we find that  $\rho\sqrt\sigma \sim 0.6$. This gives an
independent evidence that exactly for the lowest mass state the flux tube
picture must necessarily break down, and that some kind of self-interaction of
the flux tube with itself must be taken into account.

The flux tube picture which emerges from the above considerations is strongly
supported by the comparison, made in tab.9, with 
the  SU(2) spectra in (2+1)
dimensions (data taken from ref.\cite{tep1,tepmor}). 
It is in fact very interesting to see that, again, the major disagreement 
between the two spectra is in the lowest state, which is the one with the
smallest radius, while states with higher mass 
(and larger size) show a remarkable agreement, not only in the pattern of the
various states, but also in their numerical value.
As mentioned above, for the lowest state, it is the flux picture itself 
that breaks down  and
the gauge group becomes important. On the contrary for the higher states the
choice of the gauge group seems to be ininfluent.
This observation supports the 
idea, discussed in the introduction, that these massive spectra share 
some sort of ``super-universality''  as a consequence of the fact
that their general features are all
described by the same flux tube model. This ``super-universality'' then breaks
down exactly when the flux tube picture itself looses its validity.
It would be very interesting to test this conjecture with some other 
glueball spectra in (2+1) dimension.

The overall impression from this analysis is that in general, the idea that the
glueball spectrum could be described by some type of flux tube model is
strongly supported by our results, and in particular by the fact that 
 the higher states of the spectrum seems to be independent from the gauge
 group. The IP model, which is the simplest possible realization of such a flux
 tube,  seems  able to catch (at least at a qualitative level)
some of the relevant features
of the glueball spectrum. However it has also some serious drawbacks which 
cannot be solved by simply tuning the
$c_0$ and $f$ parameters
and certainly require some improvements and new ideas.

\section{Conclusions}
 For the first time we present a detailed study of the glueball-spectrum of the
 Ising-gauge theory in 2+1 dimensions. The simulations have been performed 
 with a local demon-algorithm in Multi-spin-coding technique.  In order to 
 obtain accurate results for the spectrum, and in particular to obtain 
 results for secondary states in a given channel we used the variational 
 technique
 discussed in refs. \cite{kro,lw}. 
 We obtained  reliable estimates for 11 masses in 
 total for 5 channels. 
 We compared our Montecarlo results for the $0^+$ state with those of biased 
 inhomogeneous differential approximants of a low temperature expansion \cite{at}. 
 The results are consistent, but certainly our MC result is more accurate 
 for $\beta$ close to criticality. 
 We studied the scaling of the masses with the square  root of 
 the string tension and
 extrapolated their values to the continuum limit. These results 
 were then compared with those obtained in perturbation theory 
 for the (2+1) dimensional  $\phi^4$ model. 
 We found an excellent agreement, and our error-bars are considerably 
 smaller than those of the estimates obtained with perturbation theory.  
 In order to test the flux-tube picture of the glueball discussed in 
 the introduction we compared our Montecarlo results for the Ising model with 
 masses obtained from the IP model and Montecarlo results for the SU(2) gauge 
 theory \cite{ip,tep1}. 
 We found a remarkable agreement between the SU(2) and the Ising
 spectra except for the lowest state $0^+$. We also observed that 
 the IP model is able to describe (at least at a qualitative level) 
 some of the relevant features of the glueball spectrum.
 These results support the idea that
 the large distance properties of any (lattice) gauge theory in the confining
 phase can be described by a very general and simple flux tube model, and that
 such a picture breaks down on scales comparable with the flux tube thickness,
 where the self-interaction of the flux tube becomes important. This scale, in
 the Ising case seems to coincide with that of the $0^+$ glueball.
 The fact of working in (2+1) dimensions allowed us to deal with a 
 simplified theoretical setting and thus to test in a rather rigorous way 
 the IP model, which is the simplest possible realization of a flux tube model.
 As mentioned above the IP model correctly describes some of the features of 
 the spectrum, however it has also some serious drawbacks which 
cannot be solved by simply tuning the various phenomenological parameters of
the model and certainly require some improvements and new ideas. 

 Finally it is important to stress that, by means of a suitable duality
transformation, all our results are also valid in
three dimensional spin Ising model. 
It would be very interesting  to see if
the knowledge of the first states of the spectrum could give us some hint 
toward a better understanding  of the three-dimensional Ising model.

\vskip 1cm
{\bf  Acknowledgements}

We thank F.Gliozzi, K.Pinn, P.Provero and S.Vinti for many helpful 
discussions. 
M. Hasenbusch expresses his gratitude
for support by the Leverhulme Trust under grant
16634-AOZ-R8 and by PPARC.

\newpage
\centerline{\bf Figure Captions}
\begin{description}
\item{\bf Fig.1} \hskip 0.5cm 
Some lattice realizations of the operators discussed in sect.~3.2~. 
To clarify the role of the various symmetries, for the $2^-$ and 1/3 channels
we have shown  respectively two and four different realizations. 
For the $0^+$, $0^-$ and $2^+$ only the simplest possible realizations are
shown.

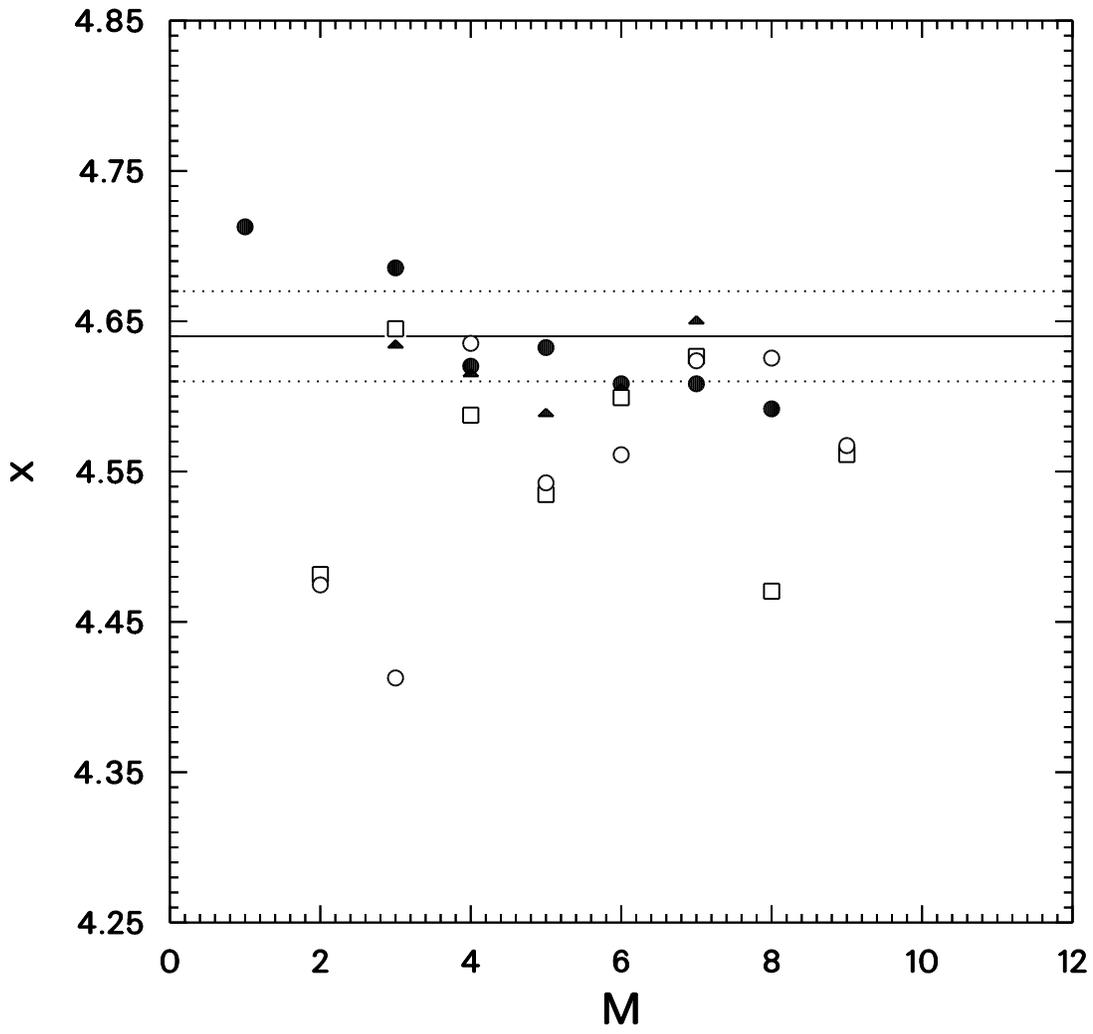
\item{\bf Fig.2} \hskip 0.5cm 
Approximants of $\xi$ at $\beta=0.75632$  for a 
large range of [K/L;M]. 
The continuum line and the two dotted lines denote the mean value and the error
of our montecarlo estimate :$\xi=4.64(3)$. The symbols correspond to different
values of K. Open circles: K=4, squares: K=5, black circles: K=6, triangles:
K=7.

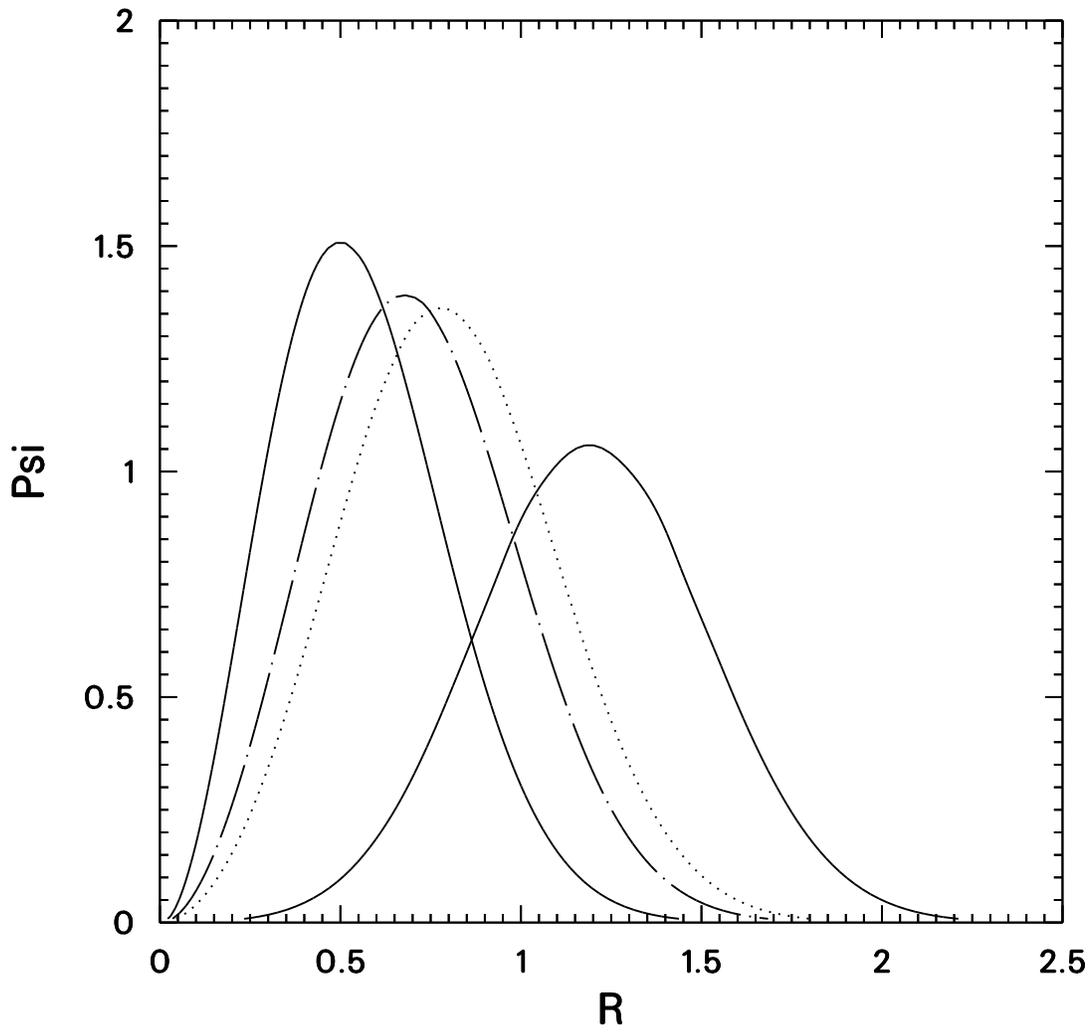
\item{\bf Fig.3} \hskip 0.5cm 
Eigenfunctions  obtained with the ``ideal string'' IP model ($c_0=0$,
 $f=\infty$)    for the lowest excitation in each channel.
The radial coordinate is measured in units 
of $1/\sqrt\sigma$. The four curves correspond, 
in order of increasing radius, to
the $0^+$, $2^\pm$, $3^\pm$ and $0^-$ states.

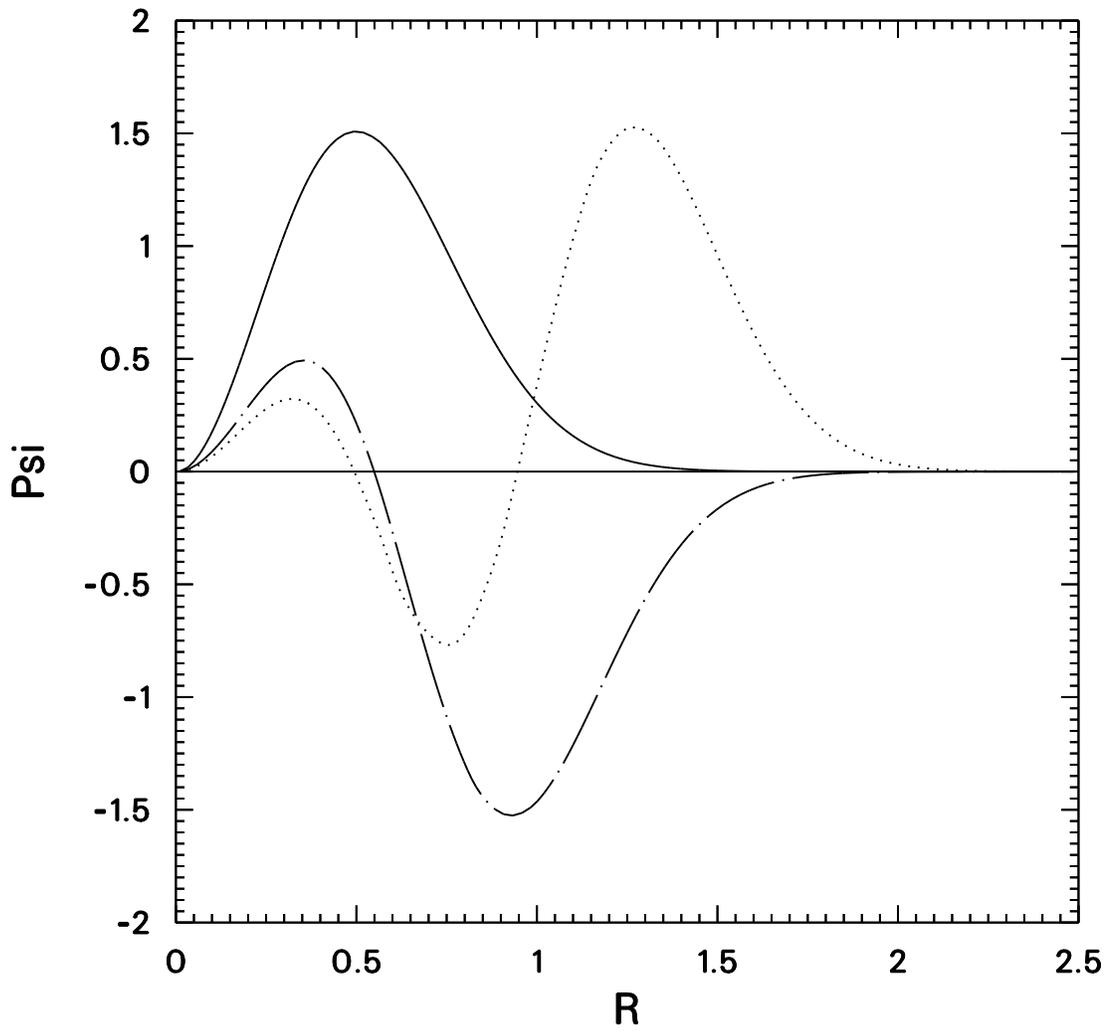
\item{\bf Fig.4} \hskip 0.5cm 
Same as fig.3, but for the ground state and the two first radial excitations 
of the $0^+$ channel.

\end{description}

\end{document}